\documentstyle[12pt,aasms4]{article}

\begin{document}

\title{Young Supernova Remnants and the Knee in the Cosmic Ray Sectrum}
\author{Anatoly Erlykin}
\affil{Physics Department, Durham University, Durham DH1 3LE, UK}
\affil{Lebedev Physical Institute, Moscow 117924, Russia}
\author{Tadeusz Wibig}
\affil{Physics Dept., University of \L \'{o}d\'{z};\\ The Andrzej So\l tan Institute for Nuclear Studies, Uniwersytecka 5,
90-950 \L \'{o}d\'{z}, Poland.}
\author{Arnold W. Wolfendale}
\affil{Physics Department, Durham University, Durham DH1 3LE, UK}
\authoremail{wibig@zpk.u.lodz.pl}

\begin{abstract}
It has recently been suggested that neutron stars inside the shells of young supernova remnants (SNR) are the sources of PeV cosmic rays and that the interaction of the particles with the radiation field in the SNR causes electron pair production, which has relevance to recent observations of 'high' positron fluxes. Furthermore, the character of the interaction is such that the well-known knee in the cosmic ray energy spectrum can be explained. Our examination of the mechanism leads us to believe that the required parameters of SN and pulses are so uncommon that the knee and positron fraction can only be explained if a single, local and recent SN -- and associated pulsar -- are concerned.
\end{abstract}


Although it is over 50 years since the 'Knee' in the cosmic ray (CR) spectrum, at about 3 PeV, was discovered, its origin is still the subject of controversy.  Many have considered that it is simply due to Galactic Diffusion, the entrapment of CR by Galactic Magnetic fields becoming increasingly inefficient above this energy.  We ourselves, however, favour a 'single source model' in which a single, recent local supernova (SN) is responsible (e.g. Erlykin and Wolfendale, 1997, 2001).  We have argued that the knee is too sharp to allow Galactic Diffusion to work.  This view has relevance to the recent work by Hu et al (2009), (to be referred to as I) which includes the observation of a sharp knee in the all-particle spectrum measured by the impressive Tibet AS-$\gamma$ array (Amenomori et al, 2008).

The advance made by Hu et al. (2009) is their proposal that pulsars (P) close to their parent young supernovae (SN) can give a spectrum of the appropriate shape, viz a sharp knee.  The mechanism involved is nucleus-SN optical radiation interactions. A bonus is an explanation of recent excesses of electron/positron fluxes.  

Hu et al. choose appropriate parameters and conclude that there are 3 possibilities: (1) all the sources (to be designated P, SN) are 'standard', (2) the average effect is equivalent to using  one set of parameters or (3) one single nearby source dominates the observed fluxes of CR.  In what follows we endeavour to determine which, if any, of these possibilities is valid.
Insofar as they, and we, incline to the view that helium nuclei predominate at the knee, and, indeed the knee is in the helium component itself, we concentrate on this component.  In I the parameters used (needed) are: an effective black body radiation temperature of 7000K, a period for acceleration of $\tau$ = 0.19 y and $nc\tau = 12.9 \times 10^{29}$ cm$^{-2}$ (where $n$ is the number of photons per cm$^{3}$  in the interaction region).
We will look at various parameters, in turn:
\begin{itemize}
\item[]The 'sharpness' of the predicted helium spectrum. 
\item[]
The characteristics of pulsars.
\item[]
The characteristics of young SN.
\end{itemize}
In a number of papers we have examined the sharpness of the knee from the published spectra in many experiments, most recently in Erlykin and Wolfendale (2001); to be referred to as II.  We have defined 'sharpness', $S$, as the second differential of the logarithm of the intensity with respect to the logarithm of the energy.  In II we pointed out that GM prediction was $S=0.3$; anything above this is regarded as needing a contribution from a sharply peaked (in log intensity versus log energy) spectrum, delivered from a single source.
In II the adopted model gave $S\approx -1.8$ for the all particle spectrum, using $\Delta \log E = 0.2$ in the derivation of $S$.  The adopted model spectrum for helium (the preferred nucleus at the knee) had a very sharp cut-off in intensity, in fact, an unnecessary feature.   In I the helium component itself has $S = -3.1$ and the all -- particle spectrum has $S = -1.7$, (the '3.1' is diluted by the smoothly varying intensities for the other mass components).  The spectral knee shown in  I for the Tibet-III spectrum is similarly sharp.

The near equality of the S-values from I and II means to us that the single source model (II) gives an all particle spectrum in good agreement with that predicted in I.

It remains to examine the reasonableness of the new model involving PSN (I) and, in particular to see whether 'standard' sources of the type specified are likely to occur.

We start with the pulsars.  Millisecond pulsars are involved, following the work of Gaisser et al. (1999).  Such pulsars (period $\lesssim$10 ms) and having high magnetic fields ($\gtrsim$10$^{12}$ G) are required,. Here we encounter the first snag, however: it is apparent that such pulsars are very rare.  Millisecond pulsars themselves account for about 3\% of all pulsars but most of these are low magnetic field 'spun-up' objects and the number of potentially useful millisecond pulsars is less than 1\%.  This is too few to give the whole of the PeV CR spectrum.  The recently discovered 'magnetastars' (Kasen and Bildsten, 2010), may provide the answer, however. These objects appear to have periods in the range 2-20ms, magnetic fields $\sim 5 \times 10^{14}$ G, and luminosities above about 10$^{48}$ erg s$^{-1}$. The bulk of their energy loss occurs over days to weeks. Their frequency may be as high as a few \%.

Turning to the associated young SN, we agree that of order 1 eV photons are common and that $nc\tau \sim 10^{30}$ cm${-2}$ is both needed and available.  However, it appears that the condition of a near constant photon energy during the interaction period ($y \sim 0.2\gamma$) is not (often?) met.  The brightness, too, is, hardly surprisingly, variable, depending on such parameters as progenitor radius, the ejected mass and the explosion energy (e.g. Young, 2004).  The spectra of different SN differ somewhat from one to another, too, and are not black body, some also emit copious fluxes of soft X-rays, eg $3\times 10^{46}$ erg over 1000s for red supergiants (Nakar and Sari, 2010) which would interact with lower energy nuclei and cause complications. 

Of the above, the time dependence of the mean photon energy is the most serious and this needs further examination. 3D radioactive transfer calculations have been made by Kasen et al. (2006) and can be taken as an example.  These workers quote mean temperatures as a function of time from the explosion as follows. 

$6\times 10^4$ K (5 days), $3\times 10^4$ K (10 days) and $2\times 10^4$ K (19 days). Multiplying by the radiation intensity as a function of time yields intensity times temperature for the times listed as 6, 9 and 4 units. Thus, there is near constancy of effective intensity over a range of a factor 3 in temperature, with consequent smoothing of the knee in the derived CR spectrum.

We have examined in more detail the behaviour of the sharpness of different CR mass components effected by the e$^+$e$^-$ pair production mechanism for the model of expanding SNR when the temperature varies by a factor of a few within days. For the background photon energies of a  fraction of an eV and initial energy density of 10$^{15}$ eV/cm$^3$ we found a significant effect above 10$^{15}$ eV. The sharpness parameter values obtained form upper limits which could be reduced in a particular source by the flux of CR produced and propagated undisturbed later on, when the opaqueness of the SNR diminishes. We have observed for constant SNR with photon temperatures of 500 K and 1000 K peak values of $S$ equal to about 4 and 2.5, respectively.  The temperature change from  1000 K to  500 K was chosen to obtain the peak position for the proton component at $3\times 10^{15}$ eV and the peak value is $\sim 2$ there (these temperatures are,in fact,unusually low
for a conventional SNR). The maximum of the sharpness scales with the CR particle gamma factor, is proportional to the CR particle mass number. The sharpness maximum is related to the initial part of the steepening of the spectrum: when the flux is reduced by only 20\%. The significant reduction, about a factor 5-10 appears at an energy about 5 times higher than the one for maximum $S$, independently of the CR particle mass.

The sensitivity of the sharpness, $S$, to the spread of the other parameters involved needs further examination.

Other problems of the model, which are beyond computation, include the following:
\begin{itemize}
\item[i]The effect of the shock wave of the SN, propagating at $\sim 3 \times 10^4$ km/s, on the emerging optical radiation and on the rotation characteristics of the pulsar,
\item[ii]Non-isotropic interactions between SN photons, which are strongly collimated, and the Pulsar-accelerated nuclei.
\item[iii]The likelihood of the required helium nuclei dominating in the PeV region, as distinct from other nuclei.
\end{itemize}
In II an analysis was made of $S$ as a function of the standard deviation $\sigma$ of the error in the logarithm of the energy. As a check, the form of the helium energy spectrum given in I has been taken, and uncertainties in energy of magnitude $\sigma$ (in $\log E$) have been applied. Similar results to those in II were derived; these have the following $S(\sigma)$ values with respect to unity at $\sigma = 0$: 0.6 (0.1), 0.33 (0.2) and, extrapolating, 0.1 at $\sigma = 0.42$.

Returning to the $S$-value derived for Helium in I (3.0), this resulted in a fit to the data for the all-particle spectrum with adequate sharpness, as already mentioned. We consider that a Helium S-value less than 1.0 would be quite unacceptable. From the above this indicates $\sigma \lesssim$ 0.2, or, in linear scale, $\pm 60$\%.

Applying these argument to the model advanced in I we have the following. To be acceptable there can be no deviation from the adopted model parameters overall by more than about $\pm 60$\%. For a single source, already there are difficulties in that the likely SN temperature variation gives a range of about 3, ie $\approx \pm 50$\% (ie $(1+0.5)/(1-0.5)$) and the
temperatures needed are low. Nevertheless, this model is 'in with a chance'.

As applied to the origin of the knee in terms of many standard sources of the type specified in I, however, there appears to be no possibility at all. SN vary too much from one to the next and pulsars of the required characteristics are far too infrequent.

\end{document}